# Synchronized Circular Motion of Optically Confined Marangoni Microswimmers

Sabera M. Borno[1,2], Robin Khisa[3], Israt H. Zarin[4], Md Hasan Mahmud[4], Nicholas D. Brubaker[5], Ryan C. Hayward[6], Nabila Tanjeem [6,7,*]

[1]Department of Physics, University of California, Merced, Merced, California, 95343, United States.

[2]Department of Physics, Shahjalal University of Science and Technology, Sylhet, Bangladesh.

[3]Dept. of Electrical & Electronics Engineering, Noakhali Science & Technology University, Noakhali, Bangladesh

[4]Department of Physics, University of Chittagong, Chittagong, Bangladesh

[5]Department of Mathematics, California State University, Fullerton, Fullerton, California, 92831, United States.

[6]Department of Chemical and Biological Engineering, University of Colorado, Boulder, Colorado 80303, United States.

[7]Department of Physics, California State University, Fullerton, Fullerton, California, 92831, United States.

* Corresponding author: ntanjeem@fullerton.edu

**Abstract:** Understanding the collective actuation of microscopic structures driven by external fields can lead to the development of next-generation autonomous machines. With this goal in mind, we investigated light-induced collective motion of thermocapillary microswimmers at the air-water interface. We found that Marangoni forces, which lead to long-ranged repulsive interparticle interactions, can cause microswimmers to synchronize their circular motion in a collective chase mode that resembles predator-prey behavior often observed in nature. We examined different degrees of confinement in small systems containing 2 – 6 particles of different individual swimming velocities and shapes. Thanks to the strong repulsive interactions between particles, a sustained chasing mode was observed for particle packing fractions above a critical value of 0.25. At lower packing fractions, swimmers transition between chasing, bouncing, and intermittent pausing, likely due to time-varying activity levels. Additionally, we report that a new synchronized mode can be introduced by incorporating chirality in particle shapes, where the microswimmers collectively reverse the direction of their circular motion periodically. Our results point to a simple but powerful mechanism of obtaining collective synchronization in synthetic confined systems where particles are designed with different shapes and activity levels.

# Introduction

Systems of self-propelling particles and their collective behavior, as observed in flocks of birds, schools of fish, swarms of ants, and groups of bacteria, have received significant attention in recent years. [1–7] An intriguing aspect of self-propelling entities is the non-reciprocal interaction that gives rise to predator-prey-like behavior.[8] For instance, when starling flocks encounter a predatory threat such as a peregrine falcon, they exhibit a wide array of collective escape patterns.[9] The significant features of predator-prey collective dynamics have been reproduced across multiple length scales in biomimetic synthetic materials. For example, non-reciprocal interactions have been directly programmed into mesoscale robots to obtain valuable insights into out-of-equilibrium phases.[10] At the microscale, predator-prey behavior has been engineered using oil droplets[11,12] and chemically active colloids.[13] Non-reciprocal interactions between active droplets have been predicted to give rise to different modes of collective motion—i.e., chasing, bouncing, and pausing.[14] Because of the rising development in such synthetic systems, there is now a growing need for understanding their collective dynamics for different experimentally relevant parameters, such as the size, shape, activity levels, and confinement.

Actuation of microscale objects driven by Marangoni forces is an area that has given rise to a wide range of intriguing phenomena in recent years. The modulation of thermal or chemical gradients at the interface of two media allowed for the design of camphor surfers,[15–17] colloidal scale microswimmers, [18–20] micro-robots,[21–24] and logic gates.[25] Swimmers with asymmetric shapes have been used to demonstrate chiral trajectories that arise from the coupling of translational and rotational motion.[26–29] Beyond single microswimmers, the collective dynamics of multiple Marangoni swimmers have been explored both experimentally and via simulations.[30–35] The advantage of Marangoni swimmers driven by photothermal fields derives from the ability to modulate their collective dynamics on demand. Despite the recent progress, the simple yet intriguing predator-prey dynamics have not been realized using Marangoni swimmers. Besides, it remains unknown how the local environment of individual swimmers, such as shape and swimming velocity, may affect the collective dynamics.

In this report, we demonstrate that long-ranged thermocapillary interactions can give rise to predator-prey-like behavior of Marangoni microswimmers in an optically defined confined region. The Marangoni swimmers used in our experiment are photothermal particles that generate heat under light illumination and thereby modulate the surface tension of the surrounding air-water interface. Prior studies have shown that two adjacent thermocapillary swimmers experience a long-ranged repulsive force arising from their temperature fields.[25,36] In this work, we find that the repulsive thermal interactions can synchronize the collective motion of up to six particles moving along the boundary of an optically defined circular region. We examined the collective dynamics for different degrees of confinement, defined by $\phi = \frac{Na}{\pi R}$, where $R$ and $a$ are the radii of the confinement and the microswimmer, respectively, and $N$ is the number of swimmers. In the limit $R >> a$, this represents the linear packing fraction of particles along the ring at the boundary of the confining circle. We find that the particles synchronize into a "chase" mode for packing fractions higher than 0.25, where all particles chase each other while maintaining a constant interparticle separation distance. The particles exhibit weaker synchronization for lower packing fractions (0.10-0.25), and switch between chasing, pausing, and bouncing modes. Using an inertial model of coupled microswimmers, we explain that the synchronized chase mode in circular confinement arises from the strong repulsive interactions of the two nearest neighbors that help sustain constant velocities and interparticle separation distances over time. For particles with elliptical or chiral shapes, as well as mixtures of different shapes, we observed consistent tendencies for increased synchronization in smaller confinement. Interestingly, we found that groups of chiral swimmers exhibit temporary and periodic collective reversing in their circular swimming trajectories that likely arises from the coupling between their rotational and translational motions. These results have important implications for synthetic platforms that apply collective dynamics for active transport, micromanipulation, and mixing at the microscale.

## Experimental Section

### 1. Fabrication of photothermal microswimmers

The photothermal microwsimmers used in our experiment are hydrogel nanocomposite disks (HND) fabricated using a method published in our prior work.[25,37,38] Briefly, we synthesized a copolymer, poly(diethylacrylamide-co-N-(4-benzoylphenyl)-acrylamide-coacrylic acid) and mixed it with gold salt ($AuCl_3 \cdot 3H_2O$). Then, we applied a two-step photolithographic method– (i) to crosslink the polymer with UV light (365 nm) and (ii) to form gold nanoparticles via photocatalytic reduction inside the crosslinked polymer using blue light (400 nm). Following exposure, the uncrosslinked polymer was removed using a developer solution and the HNDs were released in ultrapure water. The microswimmers used in the experiment were fabricated with diameters in the range of $d = 2a = 300 - 400$ μm.

### 2. Characterization of light-induced swimmer dynamics

We studied the behavior of the Marangoni swimmers by placing them at the air-water interface of a sterile petri dish (Fisher Scientific, FB0875713A) full of fresh ultrapure water.[25,38] Microswimmer dynamics were observed using bright-field illumination in an inverted microscope (Nikon ECLIPSE Ti) equipped with a 10x objective lens. The light irradiation experiments were performed using a white light source (Lumencor Spectra light-emitting diode) and a digital micromirror device (DMD). The non-uniform brightness profile was implemented by applying a grayscale pattern and then assigning each pixel on the projected image a probability of being "ON" that is proportional to the gray value of that pixel (Figure S1).

### 3. Tracking swimmer motion

Particle positions were tracked from different movie frames using the Python OpenCV package. Particles with uniform circular shapes were detected using the Hough transform method. The estimated range of the particle radii and edge detection parameters was optimized to detect the particle positions accurately. To track the particle positions over time, each particle was linked to the particle found in the subsequent frames within a small search window. The chiral, elliptical, and other shapes were detected using the template matching functions of OpenCV. We developed an algorithm that rotates a template shape within a predefined range of angles and compares it to the image of a particle to identify the best matching result, which estimates the value of the rotation angle $\varphi$.

## Results and Discussion

### 1. Circular motion of a thermocapillary microswimmer in optical confinement

The microswimmers used in our experiment are hydrogel nanocomposite disks (HND) with gold nanoparticles embedded in them (diameter $d = 2a = 300$ μm $- 400$ μm) fabricated using a previously published method.[20,29,30] Under light irradiation, the photothermal response of the gold nanoparticles generates a temperature and surface tension gradient around each microswimmer at the air-water interface. The optical confinement is realized by defining a circular region of low light intensity ('darker') inside a high intensity ('brighter') ring (as shown in Figure 1a, b). When a particle approaches the boundary between the brighter and the darker region, it gets pushed toward the darker region with a lower temperature and higher surface tension. The Marangoni force acting on the particle can be expressed as $F \approx \gamma_T a (T_{bright} - T_{dark})$, where $\gamma_T$ is the variation in surface tension with temperature, $T_{bright}$ and $T_{dark}$ are the temperatures of the brighter and the darker region, respectively. The temperature increase around a particle can be expressed as $\Delta T \approx \frac{Q}{2\pi k r}$, where $Q$ is the total heat generation by the particle which is proportional to the light intensity $I$ ($Q \propto I$), $k$ is the thermal conductivity of water, and $r$ is the distance from the particle center. The input light intensity in the experiment was maintained at a constant value of 1.0 W/cm². The darker and brighter regions were generated by projecting grayscale images onto a digital micromirror device.

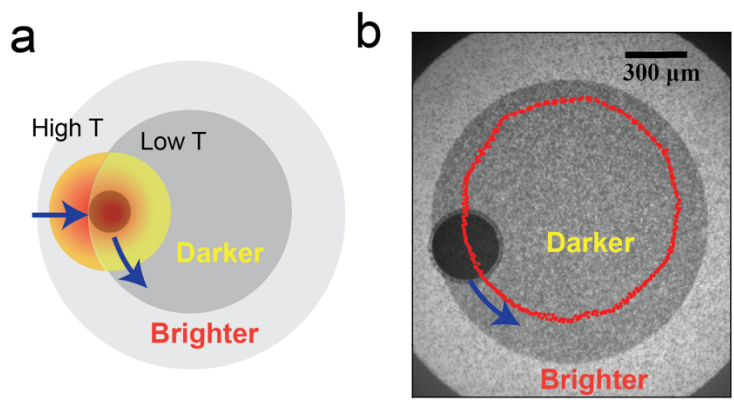

Figure 1 Circular motion of a microswimmer under optical confinement. (a) An illustration of a thermocapillary microswimmer (represented by the red circular particle) confined in the boundary between a brighter (high temperature) and a darker (low temperature) region. The temperature difference at the boundary causes a surface tension gradient, resulting a Marangoni force on the particle that keeps it on a circular trajectory. (b) Experimental demonstration of a particle (black circle) swimming in a circular trajectory defined by the light pattern. The red dots show the trajectory of the swimmer, and the blue arrow shows the counterclockwise direction of swimming.

Once placed in an optically confined region, we find that HNDs can spontaneously break symmetry, although a uniform light intensity is applied across the confined zone. This autophoretic swimming resembles the motion of symmetric camphor swimmers caused by advection and diffusion.[15,16,39] The fluid flow around a moving particle advects heated fluid from in front of the particle to behind it, causing a lower temperature and higher surface tension at the front of the particle, thereby pulling it forward. When confined inside a circular region that is darker than the background, the presence of a radially inward Marangoni force makes the circular orbit a stable trajectory. As a result, we observe the circular motion of the particle along the boundary of the confinement, as shown by the particle trajectory in Figure 1b. Although we found that the swimming velocity of a particle increases with light intensity, we measured a wide distribution of particle velocities across different samples for a given light intensity (see SI section 2 for detailed characterization). The difference in swimming velocities can be attributed to a variety of factors - variation in surface-active impurities that can cause solutocapillary counterflow to inhibit thermocapillary forces, contact-line pinning, and non-uniform gold nanoparticle formation across the particle surface.[18,20,40] Figure 1b shows the experimental observation of a microswimmer demonstrating a counterclockwise circular motion in a confined region of radius $R = 4a$ with an angular speed of 90 degrees per second (15 rpm).

## 2. Synchronization of two confined microswimmers

When two microswimmers are placed in an illuminated region in the vicinity of each other, they experience a long-ranged repulsive force. The repulsive Marangoni force on a particle exerted by its neighbor is $F_{\text{repulsion}} \approx \frac{\gamma_T Q a^2}{2\pi k r^2}$, as estimated and experimentally characterized in our prior work.[25,38] We find that this repulsive interaction causes two confined particles to synchronize their circular motion in a number of different modes - chase, bounce, and a combination of multiple modes. First, we show evidence that synchronization can occur for a pair of swimmers with different individual velocities. Second, we find that the coupled swimming can transition from one mode to another even during the same experiment, especially when the particles are confined in an area a few times larger than the particle diameter.

**Synchronization of a microswimmer pair with different individual swimming velocities:** We examined the synchronization of two microswimmers whose individual swimming velocities are different by an order of magnitude (Figure 2a). Particle 1, when swimming alone, has an angular speed of only 0.5 rpm compared to a much faster particle 2, with an angular speed of 14.5 rpm. When placed together in a confined circle of $R = 3.65\ a$, the particles synchronize their motion after 20 s (Figure 2b). At the synchronized state, they maintain an average separation distance of $918 \pm 50$ μm. The angular speed of particle 2 was found to be $8.2 \pm 2.4$ rpm, chased by particle 1 with a nearly equal speed of $8.2 \pm 1.3$ rpm. During the first 20 s before synchronization, particle 1 was stationary in a "pause" mode, while particle 2 was "bouncing". The onset of the chase synchronization was likely triggered by a sudden increase in the velocity of particle 1, as shown by the green plot in Figure 2b.

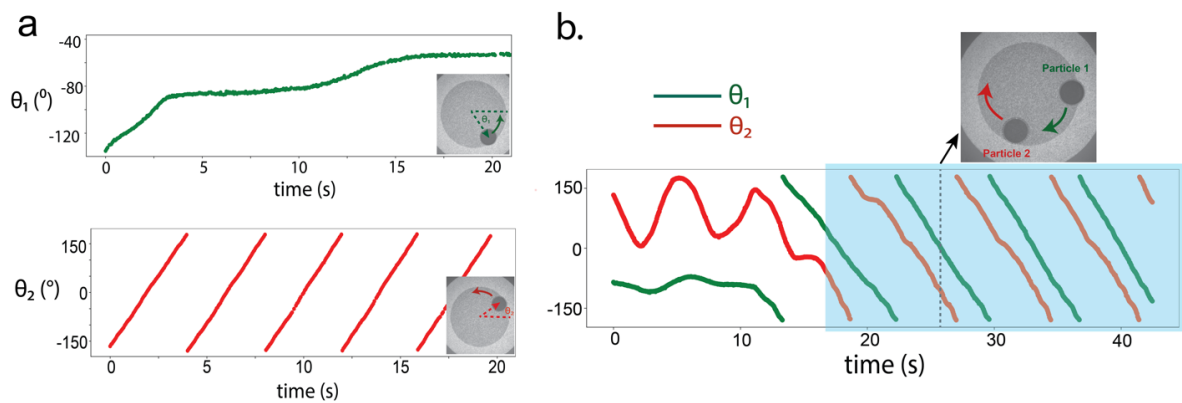

Figure 2 Synchronized chasing motion of two Marangoni microswimmers (a) Individual swimming trajectories of the swimmers when placed separately in optical confinement. $\theta_1$ and $\theta_2$ are polar angles representing the circular motions of particle 1 and particle 2, respectively ($-180° < \theta_1, \theta_2 < 180°$). (b) Swimming trajectories of particle 1 and 2 when they are placed together in a confinement. The blue shaded region shows the timespan when they synchronize their circular motion, with particle 1 chasing particle 2 at the clockwise direction.

**Transition between bouncing and chasing:** When two swimmers with similar individual swimming velocities were placed in a relatively larger confinement of $R = 4.53a$, a transition between "chase" and "bounce" mode was observed. As shown in Figure 3a and 3b. During the first 5-10 s, both particles rotate in a counterclockwise (CCW) direction, maintaining a separation distance of $1202 \pm 87$ μm and comparable velocities (particle 1:13.8 rpm, particle 2 :12.6 rpm). At about 15 s, particle 1 (green) starts slowing down, which forces particle 2 to switch its direction into a clockwise (CW) rotation. The distance between the two particles is about 727 μm when the switching occurs, which is about two times larger than the particle diameter. After 30 s, we again observe the synchronized rotation of particle 1 and particle 2, however, in a different "bouncing" mode. In this mode, particle 1 and particle 2 rotate in opposite directions (CW and CCW) and switch their swimming directions once they get close to each other (distance about 600 μm). The interparticle separation distance during the bounce mode ($1016 \pm 229$ μm) shows a larger deviation over time compared to the chase mode. The velocities of the two particles oscillate between -15 rpm and +15 rpm during bouncing, as shown in Figure 3c.

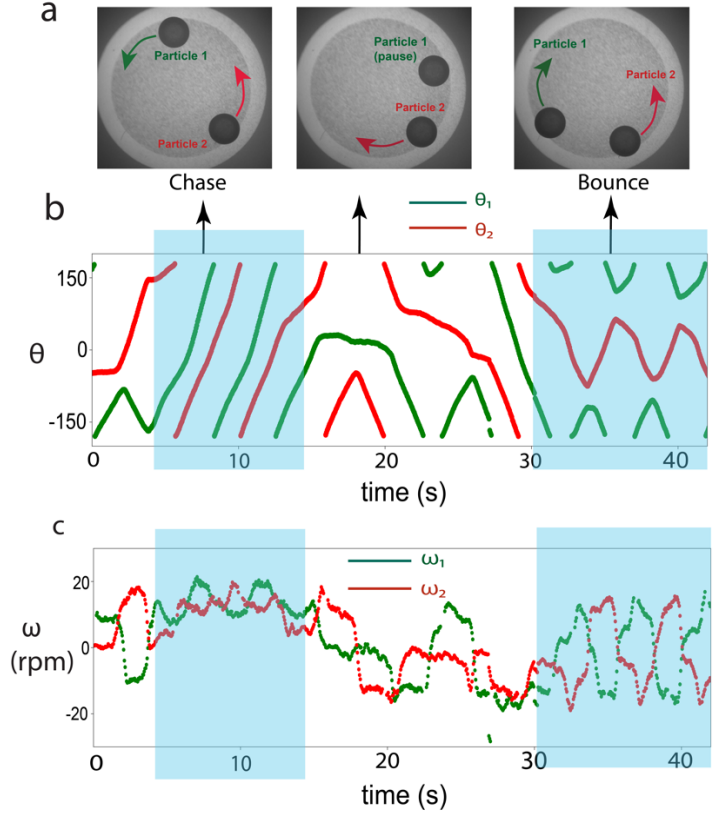

Figure 3 Synchronized motion of two microswimmers in a combination of chase, pause, and bounce mode (a) The position and swimming direction of the two particles at three different times – chasing (left), pausing (middle), and bouncing (right) (b) The angular displacements of the two particles, showing that they sustain a chase mode during 5 – 15s, and initiate the bounce mode from 30 s onward. (c) The corresponding angular speeds of the two particles during the timespan of the synchronization. The chase mode in between 5-15 s shows near-constant velocities of both particles, whereas the bounce mode after 30 s shows that the velocities oscillate between their maximum and minimum values in opposite phases. In between the chase and the bounce mode, one of the particle velocities (green, followed by red) approaches zero, which likely causes the loss of synchronization.

To explain the synchronized swimming modes, we first apply a Kuramoto model[41–43] that includes inertia and repulsive interaction between two microswimmers. The following coupled differential equations were solved for the two microswimmers with constant activity levels, $\omega_{01}$ and $\omega_{02}$.

$$m\ddot{\theta}_1 = -\eta(\dot{\theta}_1 - \omega_{01}) - \frac{K}{d_{12}^2} sin(\theta_2 - \theta_1) \quad (1)$$

$$m\ddot{\theta}_2 = -\eta(\dot{\theta}_2 - \omega_{02}) - \frac{K}{d_{12}^2} sin(\theta_1 - \theta_2) \quad (2)$$

where $m$: particle mass, $\eta$: effective viscous drag coefficient, $d_{12}$: separation distance between the two particles, and $K$: coupling constant. The solutions for three different initial conditions with constant $\frac{\Delta\omega}{K} = \frac{\omega_{02}-\omega_{01}}{K}$, but different $\omega_{01}$ and $\omega_{02}$, are shown in Figure 4. We observe synchronized chase modes for $\omega_{01} \neq \omega_{02}$ (Figure 4a and 4c), when the swimmer with the higher speed becomes the chaser. The swimmer with the lower speed matches its direction of rotation to the chaser, causing both particles swim with a constant speed of $\frac{\omega_{02}+\omega_{01}}{2}$. This finding is consistent with the experimental observation mentioned above where two swimmers with individual speeds of 0.5 rpm and 14.5 rpm, swim at an average speed of 8.2 rpm during chasing. We observed the bounce mode of synchronization in the model only when the swimmers start with equal speed but approached each other from the opposite direction ($\omega_{01} = -\omega_{02}$), and when the value of the drag coefficient $\eta$ is smaller compared to the coupling constant $K$ ($\frac{\eta}{K} = 0.1$) (Figure 4b).

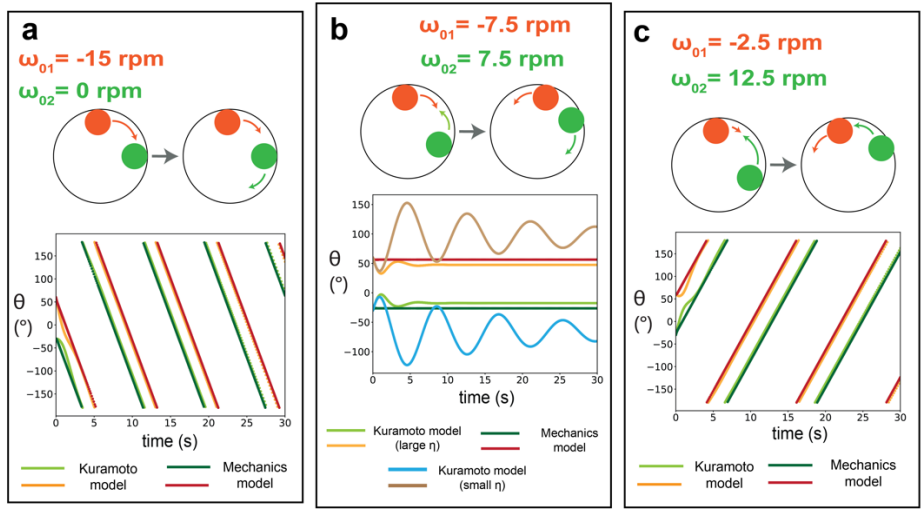

Figure 4 Results from the Kuramoto model and the Mechanics-based model of microswimmer synchronization. (a) Initially, particle 1 starts with a greater swimming velocity of 15 rpm in the CW direction and particle 2 is stationary. Particle 1 starts chasing particle 2 and both continue swimming in the CW direction. Both models can reproduce the chase synchronization. (b) Particle 1 and particle 2 approach each other from the opposite direction with angular velocities of -7.5 rpm and 7.5 rpm. The Kuramoto model with smaller drag shows the particles bounce back and forth until the oscillations get damped. The Kuramoto model with larger drag and the Mechanics-based model exhibit overdamping of the bounce mode resulting in stationary particle positions. (c) Particle 2 starts with a larger velocity in the CCW direction compared to particle 1's velocity in the CW direction. In this case, particle 2 chases particle 1 and forces it to adopt a CCW direction. Both models can reproduce the chase mode successfully.

Additionally, we apply a mechanics-based model to understand the synchronized motion in the experimentally relevant parameter regime. Note that equations (1) and (2) resemble the circular motion of two particles swimming along a trajectory with a fixed radius $R$.

$$mR\ddot{\theta}_i = \sum F_\theta = -F_{drag} + F_{repulsive} = -\eta R(\dot{\theta}_i - \omega_{0i}) + \frac{\gamma_T Q a^2}{2\pi k d_{12}^2} sgn(sin(\theta_j - \theta_i)) \qquad (3)$$

where $F_{repulsive} \cong \frac{\gamma_T Q a^2}{2\pi k d_{12}^2}$ is the Marangoni repulsive force along the azimuthal direction, the value of which has been estimated in prior experiments.[25,38] The term $sgn\ sin\ (\theta_j - \theta_i)$ was used to correctly determine the positive/negative sign of repulsive force. We find that the mechanics-based model can reproduce the chase mode successfully; however, it shows a constant angular position over time for both particles when the initial condition $\omega_{01} \cong -\omega_{02}$ is applied (Figure 4b). We can understand this result from the opposing direction of the repulsive force and the Marangoni force as the swimmers approach each other from the opposite direction. We also notice that the oscillations observed in the Kuramoto model for both small and large $\eta$ get weaker over time, indicating that the bounce mode cannot be stable when the activity levels of the particles (as quantified by $\omega_{01}$ and $\omega_{02}$) are constant.

By combining the results from the modeling and the experiment, we can conclude that the chase synchronization of two inertial microswimmers with different activity levels can be attributed to the long-range repulsive interactions between them under the assumption that their activity levels remain constant over time. However, the transition between chase, bounce, and pause mode observed in some experiments where the velocity of the swimmers has large fluctuations (as shown in Figure 3c) cannot be explained from this simplified model.

### 3. Effect of confinement on synchronized circular motion

We study the impact of the circular confinement on the synchronization modes of the microswimmers for different packing fractions ($\phi = \frac{Na}{\pi R}$) implemented with different values of $N$ and $R$. The two experiments mentioned in section 2 had packing fractions of 0.17 (Figure 2) and 0.14 (Figure 3), respectively. To quantify the degree of synchronization, we calculated the average separation distances between neighboring particles and their standard deviations over the duration of data acquisition. A larger deviation would indicate a combination of bouncing and chasing mode (hence, a lower degree of synchronization), while a smaller deviation would indicate a consistent chasing motion (a higher degree of synchronization). As shown in the experimental results of Figure 5a and 5b, the average distances between the nearest neighbors and their variances decrease as the packing fraction increases from 0.18 to 0.38. Additionally, the synchronized "chase" mode replaces the hybrid "chase/bounce" modes for packing fractions larger than 0.25 (Figure 5b and S3). The four sets of experimental data (denoted by the four different colors) were performed for $N = 3$ and $N = 4$ particles for different confinement conditions, $R$. The angular trajectories of the particles are shown in Figure S3.

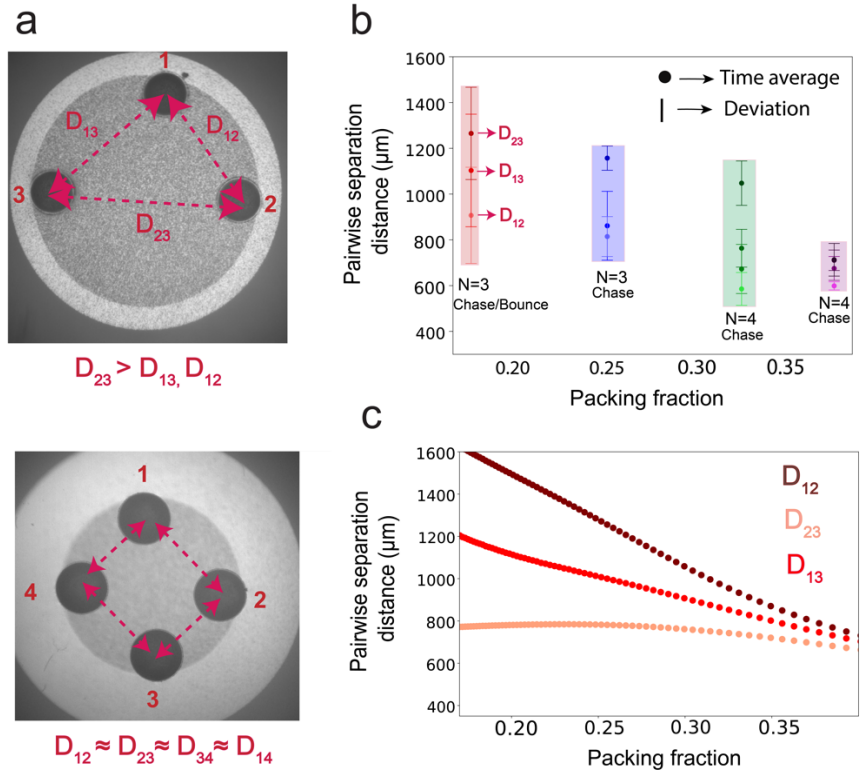

Figure 5 Effect of confinement on the synchronized mode of the microswimmers. (a) Top: three microswimmers (N=3) in a confined zone, the nearest neighbor separation distances are labeled as $D_{12}, D_{13}$, and $D_{23}$. Bottom: Four microswimmers (N=4) are in a confined zone, the interparticle separation distances between nearest neighbors are uniform due to the strong confinement. (b) Experiment results: the time average and standard deviations of the nearest neighbor separation distance are plotted for systems with different packing fractions. Each of the four colors represent one experiment with N particles in a confinement of radius R. For example, in an experiment with packing fraction $\phi$ = 0.18, the three average distances $D_{12}, D_{13}$, and $D_{23}$ are shown by the three red dots and their deviations are shown by the red-colors vertical bars. An overall decrease in the average pairwise distance and their deviations are observed for higher packing fractions (c) The nearest neighbor separation distances for different values of $\phi$ achieved from the computational model, showing that that $D_{12}, D_{13}$, and $D_{23}$ become more uniform in densely packed systems.

To further understand the impact of confinement, we solved the coupled differential equations from the mechanics-based model for N= 3 particles for different confinement sizes:

$$mR\ddot{\theta}_i = -\eta R(\dot{\theta}_i - \omega_{0i}) \pm \sum_{j \neq i, j=1}^{N-1} \frac{\gamma_T Q a^2}{2\pi k d_{ij}^2} \qquad (4)$$

We generally find that the average interparticle separation distances ($D_{12}, D_{23}$, and $D_{13}$, quantified over the course of observation time) decrease for higher packing fractions (Figure 5c), consistent with the experimental observation. With the assumption of constant activity levels ($\omega_{0i}$) for all particles in the model, we observe synchronized chase modes in the model for all packing fractions. The numerical solution shows different interparticle separation distances (i.e., $D_{12} > D_{23} > D_{13}$) for lower packing fractions, which become more uniform (i.e., $D_{12} \cong D_{23} \cong D_{13}$) for packing fractions higher than 0.25. We also examined the relationship between average nearest neighbor separation distances and packing fractions for a few different conditions of individual swimmer activity levels in the model (Figure S5). We find that the pairwise separation distance plots converge at a packing fraction higher than 0.20, indicating that swimmer synchronization for smaller confinement is less sensitive to the difference in the individual swimmer's activity level. This can explain why we observe a combination of bounce/chase modes in the experiment for smaller $\phi$ but a consistent chase mode for larger $\phi$. When the packing fraction ($\phi$) is high, every particle experiences strong repulsive force from two neighbors because of the periodic boundary condition, as depicted in Figure 5a (bottom). This condition helps the individual swimmers maintain a constant speed and sustain collective synchronization in a chase mode. As a result, the chase modes observed in the experiment are better reproduced in the model for $\phi \geq 0.25$, where the difference in individual activity levels do not seem to affect the collective synchronization.

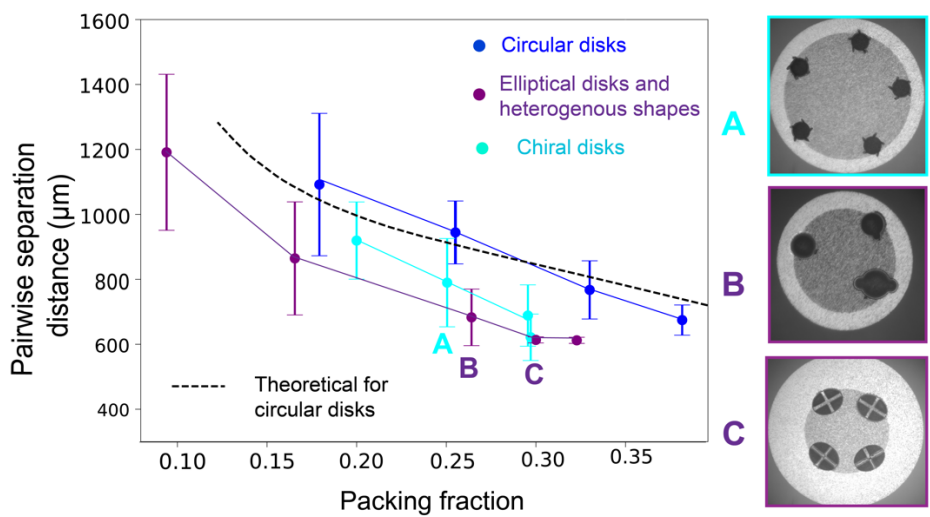

Figure 6 Effect of confinement on the synchronization of microswimmers with different shapes. Different number of swimmers with N = 3 – 6 were examined for this experiment. The average separation distances were calculated from the ensemble average of all time-averaged interparticle distances. The deviation is calculated from the average of all standard deviations over time. The images A, B, and C show the snapshots of different swimmers in synchronized chase modes under optical confinement. The theoretical plot was achieved by modeling N = 4 circular swimmers.

From the above results, we can identify a characteristic feature of confined microswimmers interacting via long-ranged repulsion – swimmers with remarkably different velocities can collectively synchronize in a packing fraction of only about 0.25 – 0.40. Next, we examine if this collective synchronization holds for swimmers with different shapes and sizes. We performed experiments with three different conditions – (i) all swimmers with chiral shapes with left- or right-handed gears (Figure 6, A in cyan color), (ii) all swimmers with elliptical shapes (Figure 6, C in purple color), and (iii) a mixture of swimmers with different sizes and shapes (Figure 6, B in purple color). The plot in Figure 6 shows how separation distance varies with packing fractions for all three swimmer sets – the cyan line represents chiral swimmers (i), the purple line represents the elliptical and heterogenous swimmers (ii and iii), and the blue line represents simple circular swimmers. We find that all swimmer groups follow a similar trend of increased synchronization at higher packing fractions regardless of their shapes. For example, the swimmers with heterogenous shapes (purple line) exhibit bounce/chase mode combinations at lower packing fractions ($\phi$ =0.10 and 0.16), but transition into pure chase modes for $\phi > 0.25$ (angular trajectories in Figure S4). The chiral swimmers also synchronize in a chase mode above $\phi = 0.20$, often accompanied by a temporary reversal in their swimming direction, which we will explain in the next section.

### 4. Collective chasing and reversing of chiral microswimmers

The shape of a microswimmer impacts its motion, as observed in chiral swimmers, where the hydrodynamic coupling between translational and rotational motions results in helical trajectories in three dimensions and circular trajectories in two dimensions.[27,44,45] We fabricated the photothermal particles with chiral shapes and observed circular trajectories at the air-water interface under light (Figure 7a). The chiral microswimmers rotate in either CW or CCW direction (as prescribed by their shapes) around an axis that intersects the center of the particles. They also swim in a circular motion, the direction of which coincides with the direction of the individual particle rotation (Figure 7a, left: CW swimmer, right: CCW swimmer). Now, we show how the circular trajectories of the individual swimmers affect the collective synchronization of a group of swimmers (N= 3 – 6) placed in an optically confined region.

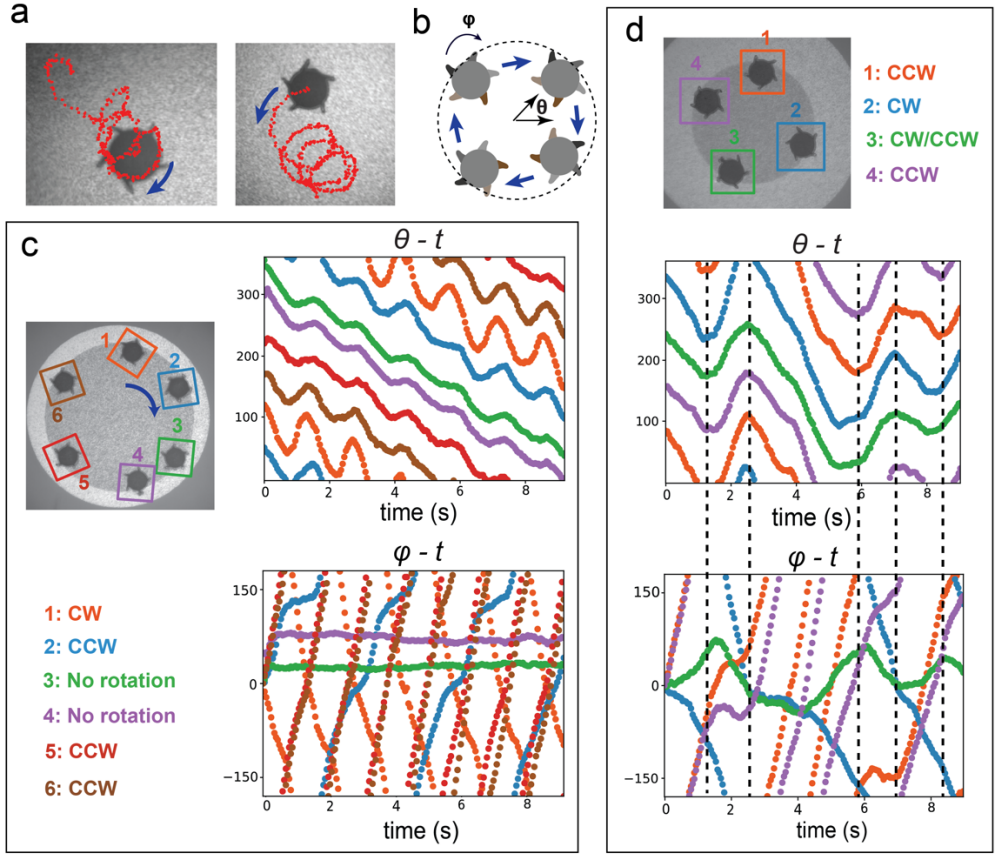

Figure 7 (a) The individual swimming trajectories of two chiral microswimmers: the left one swimming in the CW direction and the right one swimming in the CCW direction. (b) The definition of two angles, $\theta$ and $\varphi$. $\theta$ represents the angular trajectory along the boundary of the optically defined circle with radius $R$, and $\varphi$ represents the angle of rotation around the axis of each particle. (c) Collective swimming of $N = 6$ chiral microswimmers. Each particle is expressed by a number and a color, and their trajectories are shown by the $\theta - t$ and $\varphi - t$ plots. The $\theta - t$ plot shows that all particles chase each other in an overall CW direction with temporary and periodic reversals in their trajectories, denoted by the waves with small amplitudes. The $\varphi - t$ plot quantifies the angles of rotation of each particle around their own axes and shows no correlation with the $\theta - t$ plot above. (d) Collective swimming of $N = 4$ chiral microswimmers. The $\theta - t$ plot shows a collective chase synchronization whose direction reverses multiple times at t = 1 s, 2.3 s, 6 s, and 7 s, and 8.3 s denoted by the dashed vertical black lines and the waves with large amplitudes. The $\varphi - t$ plot below shows that the individual rotation of the most inactive particle #3 is affected by the collective motion of all particles and reverses five times following the pace of the collective reversal.

We observe a synchronized chase mode when N = 6 particles are placed in a confinement with a packing density of 0.3. We notice the angular trajectories of the particles exhibit sinusoidal oscillations with small amplitudes (Figure 7c, $\theta - t$ plots). The oscillations represent a temporary and periodic reversal in the direction of the collective circular motion of the swimmers. We quantified the trajectory of this circular motion using the angle $\theta$ and the angle of rotation around the particle's individual axis using the angle $\varphi$ (angles defined in Figure 7b). The $\varphi - t$ plots of Figure 7c show how the rotation of individual particles evolves over time. One particle (particle

1) rotates CW around its own axis (decreasing $\varphi$ over time), three particles (particles 2, 5, and 6) rotate CCW (increasing $\varphi$ over time), and two other particles (particles 3 and 4) exhibit negligible rotation (constant $\varphi$). The clockwise and counterclockwise directions of rotation are designed by the left-handed or right-handed gears during particle fabrication. However, some particles, such as particles 3 and 4, remain as inactive rotors, likely due to impurities on the particle surface. The major finding of this experiment is that the collective chase mode is sustained over time ($\theta$ decreases for all particles) regardless of the individual rotor orientations ($\varphi$). There are subtle differences in the angular trajectories of individual particles ($\theta - t$ plots). For example, particle 1 (orange color) exhibits the highest wave amplitude in its $\theta$ trajectory, showing that its reversal is the most significant among all particles. The reversal is likely caused by the particle's own circular trajectory as prescribed by the chiral shape. The trajectory of particle 1($\theta - t$) is most closely followed by its nearest neighbors, particle 2 (blue) and particle 6 (brown), even though they are designed with the opposite chirality. The other three particles (particles 3, 4, and 5) follow the overall direction of chase in the CW direction, but with much weaker oscillations in their $\theta - t$ plots.

We performed another experiment with N = 4 chiral particles placed in a region with an equal packing fraction ($\phi = 0.3$) but a smaller radius ($R$). In this case, we also observe a collective reversal as shown by the oscillations in the $\theta - t$ plots (Figure 7d). The amplitudes of the oscillations are higher and more uniform across all particles, demonstrating a strong collective reversal in their chase synchronization. Another interesting behavior is observed in the $\varphi - t$ plots, where the inactive particle (#3, which has very weak rotation about its own axis), starts to rotate in either CW or CCW direction (green curve on the $\varphi - t$ plots, Figure 7d). This particle switches its direction of rotation five times, exhibiting a noticeable correlation with the overall trend of the collective reversal ($\theta - t$ plots), which takes place five times during the observation period. Although the packing fraction is equal to the earlier case of N = 6 particles, the stronger interactions between the second nearest neighbor particles, such as between particle #1 and #3, and particle #2 and #4 (Figure 7d) could contribute to the stronger collective reversal as well as the rotation of the seemingly inactive rotor (#3).

Taken together, the behavior of the chiral microswimmers points to their potential for diversifying the collective synchronization modes. Since the long-ranged thermal repulsion dictates the collective motion of the particles, they all chase and reverse together even though swimmers with different chirality are present in the mixture. Additionally, in the case of a smaller confinement zone where interactions between second nearest neighbors are prominent, the rotational motion of an inactive particle can be influenced. To our knowledge, this mode of synchronization has not been observed before and could inspire future in-depth investigation on the role of chirality on collective dynamics.

## Conclusions

In conclusion, we have shown that the long-ranged thermal repulsion between Marangoni microswimmers can give rise to synchronized collective motion within an optically defined confined region. The repulsive interactions between photothermal particles at the air-water interface have been characterized in prior work and are known to cause phase synchronization in oscillators and instability in Marangoni optical traps.[25,38] In this work, we show that the repulsion can synchronize the motion of multiple particles that spontaneously break symmetry and operate like microswimmers. The synchronization takes the form of a collective chase mode in strong confinement, resembling the predator-prey-like behavior observed in many natural and synthetic systems. We demonstrate that the degree of synchronization increases with higher packing fractions, and a consistent relationship between these two parameters holds for a wide variety of swimmer shapes. We developed a simple computational model that allowed us to explain this synchronization above a packing fraction of 0.25. Finally, we show that a collective reversal in the synchronized chase mode can be achieved by incorporating chirality in the microswimmer shapes.

The implication of our experimental results is far-reaching. The collective motion of Camphor surfers in confinement is known to drive out-of-equilibrium crystallization and active turbulence.[30,31] Our study is the first to investigate the collective behavior in Marangoni microswimmers where the surface tension gradient is controlled by photothermal effects. The spatial modulation of light allows the photothermal system to define a microscale confinement without the need for a physical container. Besides, the long-ranged thermal repulsive interactions dominate the collective motion of the swimmers, giving rise to synchronization modes that are relatively simple and easier to replicate regardless of differences in particle size, shapes, and individual activity levels. The experimental results reveal important questions on the role of interparticle interactions that contribute to the collective motion. In a prior computational study, the formation of doublets or triplet in a confined region was predicted when purely hydrodynamic interactions between particles were considered.[46] Our experiments, supported by the numerical model, indicate that the long-ranged thermal repulsive interactions can enable synchronization with uniform interparticle spacing between particles - a feature distinct from the hydrodynamic interaction-based model. Further numerical studies could help us understand the interplay between these two types of interactions in the context of confined circular motion of microswimmers. Furthermore, it is known that the collective motion of active spinners can create edge flow in confined systems that can be applied for microscale transport.[47,48] Similar applications of our system may be possible where light-induced circular collective motion can transport objects to desired locations. As observed in chiral swimmer synchronization under strong confinement, the orientation of microscale objects enclosed by the rotating swimmers can be dynamically tuned, which can lead to the development of novel technologies for microscale manipulation and mixing.

**Acknowledgement:** We acknowledge funding support from the National Science Foundation (NSF) through the Brandeis MRSEC (DMR-2011846) for the experimental studies. We thank the Astro Bridge outreach program, led by Anowar J. Shajib at the University of Chicago, for including undergraduate students from Bangladesh who contributed to the computational work. We thank

Ankur Gupta and Hyunki Kim for helpful discussions on the project. We also acknowledge National Science Foundation (CBET-2301692) for supporting the work done at California State University, Fullerton.

Supporting Information

1. **Projection of grayscale images**

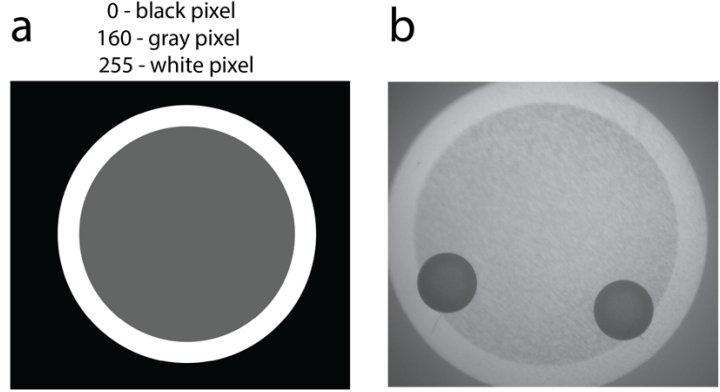

Figure S1 (a) Sample of a grayscale mask used for defining the region of optical confinement. (b) Image of the projected mask onto the microscope sample plane along with the two particles inside the confined zone.

2. **Velocity characterization of microswimmers**

We characterize the photothermal microswimmers by using two dimensionless parameters – the Marangoni number, $M = \frac{Q\gamma_T}{Ck\eta D_t}$, and the Peclet number $Pe = \frac{av}{2D_t}$. In the work of Boniface et al.[4] the relationship between these two parameters have been derived for a symmetric interfacial microswimmer:

$$Pe = \frac{M^{\frac{2}{3}}}{2\pi^{\frac{1}{3}}} \qquad (1)$$

We applied the model of autophoreic swimming developed by Boniface et al. to photothermal Marangoni microswimmers, whose temperature profile can be expressed as

$$T(r,\theta) = \frac{Q}{2\pi kr} e^{-\frac{vr}{2D_t}(1+\cos\theta)}$$

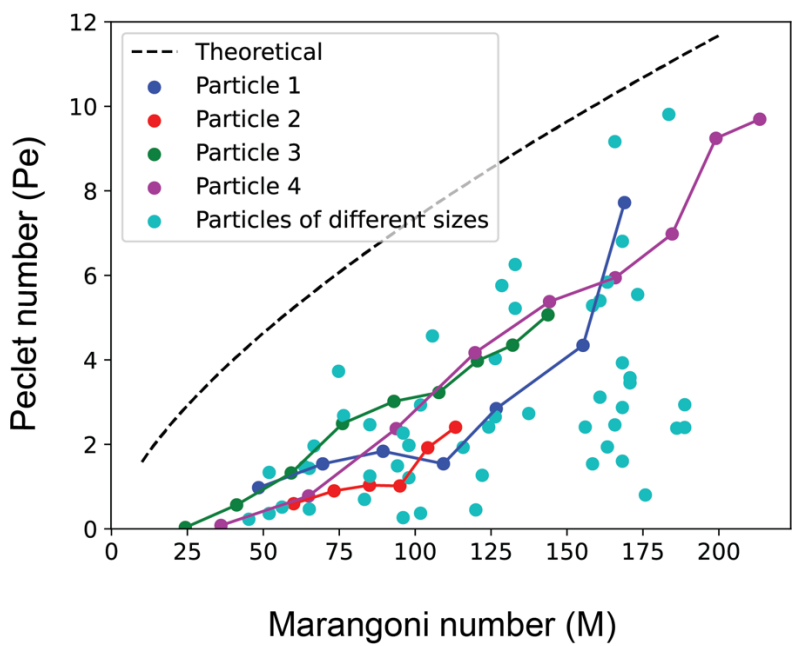

Figure S2 The Pe vs M plot of different microswimmers. The velocities of particles 1, 2, 3, and 4 were characterized for different light intensities to prepare the plot. The particles of different sizes were characterized at a constant light intensity.

Figure S2 shows Pe-M plots of microswimmers from different experimental samples (colored dots and lines) and the theoretical relationship (black dashed line) expressed by equation (1). The plot shows that particle velocities increase with incident light intensity for particles 1, 2, 3, and 4. We also find a wide distribution of velocities when different particles were examined at different environmental conditions, such as surface and water impurities (cyan points). The experimental velocities are lower than the theoretical estimates, possibly due to the contributions from the solutocapillary counterflow arising from the impurities.[5]

Below is the description of the experimental parameters used for calculating the Marangoni and the Peclet numbers:

| Parameter | Values / Description |
| --- | --- |
| $Q$: heat generated by the gold nanoparticles | $C_1 \pi a^2 I$, where $I$: incident light intensity (1 W/cm$^2$) and $C_1$: fraction of light absorbed in the confined region (0.3). |
| $\gamma_T$: variation in surface tension with temperature | 0.14 mN m$^{-1}$ K$^{-1}$ at the air/water interface |
| $C$: drag correction factor for a moving disk at an interface | $16/3$ |
| $k$: thermal conductivity of water | 0.6 Wm$^{-1}$K$^{-1}$ |
| $\eta$: dynamic viscosity of water | 0.89 Pa s at 25°C |
| $D_t$: thermal diffusivity of water | 0.146 mm$^2$/s |
| $v$: velocity of the microswimmers | Measured from the experiments |
| $a$: radius of the microswimmers | Particle 1: 198 µm, Particle 2: 158 µm, Particle 3: 178 µm, Particle 4: 208 µm<br><br>Particles of different sizes: 125 µm – 224 µm |

3. **Trajectories of N=3 and N=4 microswimmers with uniform circular shapes**

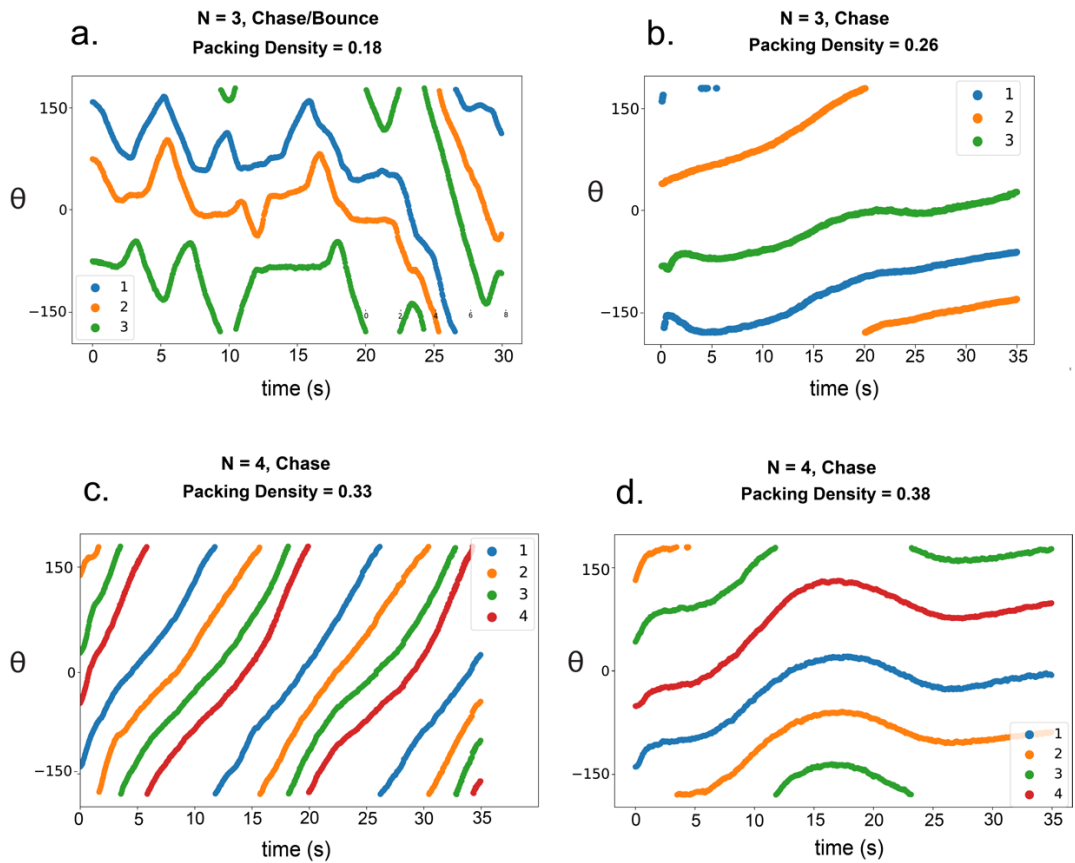

Figure S3 The angular trajectories, $\theta$ (degrees) – $t$ plots obtained from four different microswimmer experiments shown in a, b, c, and d. The swimmers show a combination of bounce, chase, and pause modes for a smaller packing fraction of 0.18 (a) but a synchronized chase mode at packing fractions greater than 0.26 (b, c, and d).

## 4. Trajectories of N=3 and N=4 microswimmers with heterogenous shapes

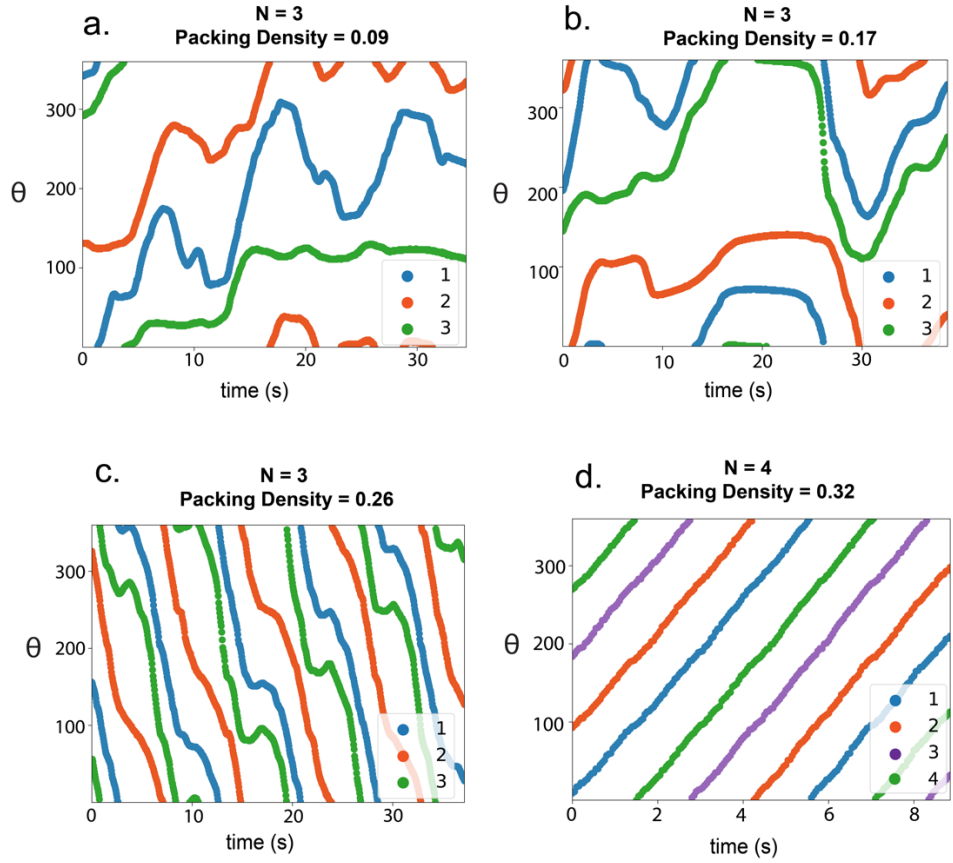

Figure S4 The angular trajectories, $\theta$ (degrees) – $t$ plots from four experiments with different microswimmer shapes shown in a, b, c, and d. The swimmers here exhibit a synchronized chase mode at packing densities higher than 0.26 (c and d).

## 5. Model of three microswimmers

Particle 1: $$mR\ddot{\theta}_1 = -\eta R(\dot{\theta}_1 - \omega_{01}) + \frac{\gamma_T Q a^2}{2\pi k d_{12}^2} - \frac{\gamma_T Q a^2}{2\pi k d_{13}^2}$$

Particle 2: $$mR\ddot{\theta}_2 = -\eta R(\dot{\theta}_2 - \omega_{02}) + \frac{\gamma_T Q a^2}{2\pi k d_{23}^2} - \frac{\gamma_T Q a^2}{2\pi k d_{12}^2}$$

Particle 3: $$mR\ddot{\theta}_3 = -\eta R(\dot{\theta}_3 - \omega_{03}) + \frac{\gamma_T Q a^2}{2\pi k d_{13}^2} - \frac{\gamma_T Q a^2}{2\pi k d_{23}^2}$$

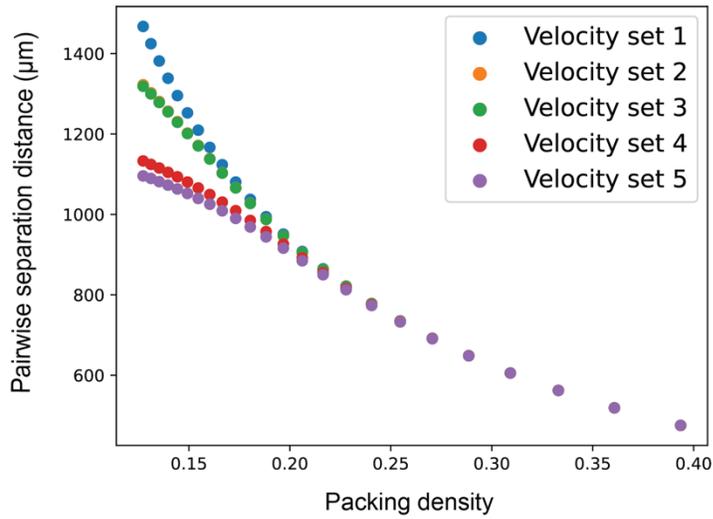

Figure S5 The average pairwise separation distances (the mean of $D_{12}$, $D_{23}$, and $D_{13}$) are plotted against packing densities for different combinations of particle velocities. The plot shows that all average distances converge above a packing density of about 0.25, hence the synchronization becomes independent of individual particle velocities.

Initial velocity conditions used in Figure S5:

|  | Particle 1 ($\omega_{01}$) | Particle 2 ($\omega_{02}$) | Particle 3 ($\omega_{03}$) |
| --- | --- | --- | --- |
| Velocity set 1 (degrees/s) | 25 | 25 | 25 |
| Velocity set 2 (degrees/s) | 25 | -25 | 25 |
| Velocity set 3 (degrees/s) | 25 | -25 | -25 |
| Velocity set 4 (degrees/s) | 25 | -100 | 25 |
| Velocity set 5 (degrees/s) | 25 | -134 | 60 |